\documentclass[10pt,twocolumn]{article}
\usepackage{graphics,graphicx,epsfig}
\usepackage{amssymb,amsmath,amsfonts}
\usepackage{ifthen}
\usepackage{epstopdf}

\newcommand{\hc}{\text{ h.c. }}

\newcommand{\lsim}{\,\raise.3ex\hbox{$<$\kern-.75em\lower1ex\hbox{$\sim$}}\,}
\newcommand{\gsim}{\,\raise.3ex\hbox{$>$\kern-.75em\lower1ex\hbox{$\sim$}}\,}

\newcommand{\LL}{\mathcal{L}}

\newcommand{\OO}{\mathcal{O}}

\newcommand{\fb}{\text{ fb}}

\newcommand{\MET}{E_T\hspace{-0.185in}/\hspace{0.075in}}

\newcommand{\GeV}{\text{ GeV}}

\begin{document}

\twocolumn[
\begin{flushright}
hep-ph/0605162\\
\end{flushright}
\begin{center}
{\huge Four $\tau$s at the Tevatron }
\vskip 0.3cm
{\normalsize
{\bf P. W. Graham$^1$,  A. Pierce$^{2}$ and J.G. Wacker$^1$}
\vskip 0.2cm
\begin{tabular}{cc}
$^1$ Institute for Theoretical Physics & $^2$ Jefferson Laboratory\\
Stanford University&Harvard University\\
Stanford, CA 94305&Cambridge, MA 02138 \\
\end{tabular}
\vskip .1in}
\end{center}

\vskip .5cm
\begin{abstract}
We study extensions of the Standard Model where the Higgs boson dominantly decays via a cascade
to four $\tau$ leptons, and discuss whether this decay is visible at the Tevatron.  We find that with an integrated luminosity of $6 \fb^{-1}$, there can be excesses in multi-lepton events in several channels for a Higgs boson of a mass $\approx$ 110 GeV.   
\end{abstract}
\vskip 1.0cm
]

\section{Introduction}
The Higgs sector is the one unexplored piece of the Standard Model. While it is possible that a single physical Higgs boson represents the entire picture at the TeV scale, considerations of naturalness suggest that it does not. Rather, it is likely there is new physics, and that new physics may well affect the phenomenology of the Higgs sector.  

The Supersymmetric Standard Model (SSM) is the leading candidate for new physics beyond the Standard Model, and the minimal model, the MSSM, has a pair of Higgs doublets with a very constrained set of interactions\cite{DimopoulosGeorgi}.  The Higgs quartic couplings, related to gauge couplings via supersymmetry, 
predict a Standard Model-like Higgs with a mass not far above $M_{Z^0}$ if the Higgs vacuum expectation 
value (vev) is not fine-tuned.  To get above the LEP2 limits on the Higgs mass\cite{LEPLimit}, even in 
the MSSM, the vev must be fine-tuned to be a few percent of its `natural' value.   This tension has 
motivated extensions to the  MSSM's Higgs sector to push the Higgs boson above the LEP2 bounds.   Many such 
extensions do not alter the search strategy for the lightest Higgs boson, but simply increase the Higgs mass.   However, it has recently been emphasized that some models can change the signals of the Higgs decays.
One class of models does so by having new light states that can be the dominant decay mode for the Higgs boson and can give rise to the Higgs decaying into invisible particles, four bottom quarks, or even more bizarre possibilities\cite{Dermisek,Neal}.  Here, we will attempt to further motivate and explore a related possibility.

If the mass of the Higgs boson is significantly beneath the $WW$ threshold, the smallness of the $b$ quark Yukawa coupling 
makes the Standard Model Higgs width very small.  The existence of any new light particle with $\OO(1)$ couplings to the Higgs will completely swamp the Standard Model decays of a light Higgs. 
%If non-standard Higgs 
%decays are open, the branching ratios to the standard decays are proportionally reduced, and the searches 
%based on the standard signatures become more challenging.  One should take advantage of the non-standard decays if possible.  However, 
If these decays have sufficiently unusual topologies, 
then they may not be triggered on and this potential discovery signal may go unrecorded.  It is important to identify these topologies to optimize searches for them.    

In this note we focus on a single possibility for a non-standard Higgs decay.  We consider a Higgs boson that dominantly decays 
to a pair of light pseudo-scalars (axions), kinematically 
limited to decay into a pair of tau leptons each.   
Thus, each Higgs boson decays to four tau leptons,
where each pair of tau leptons is collimated.   
We focus on this channel because it can arise in models of new physics in a straightforward way.  We will frame our 
discussion in terms of the NMSSM (
Next-to-Minimal Supersymmetric Standard Model) \cite{NMSSM,NMSSM2}, though the analysis is not
limited to this model.  Similar axions could be present in many models of new physics with extended Higgs sectors, including Little Higgs theories\cite{Kilian:2005xt}.  This channel is also of immediate interest 
because it might be possible for the Tevatron to find some hint of such 
Higgs decays through its production via glue-glue fusion, or in association with a weak gauge boson.  In addition, this 
decay chain seems like a challenging channel for the LHC, and studies beyond those extant are warranted \cite{GunionLHC}.  Finally, it is  a channel that might motivate new experimental triggers.

The organization of the paper is as follows.  In Section \ref{Sec: Axion} we explain a particular model where a light axion might arise and how it may dominate the decay width of the Higgs.  We also briefly 
review the existing limits on similar axions from colliders.  We then discuss possible potentials for observing the Higgs at the Tevatron, paying particular attention 
%\ref{Sec: Higgstrahlung}, we discuss the possibility of observing new physics at the Tevatron by observing events where the Higgs is produced in association with a weak gauge boson. In Section \ref{Sec: Four tau} we present a 
%searching for a 
to the case where the Higgs is produced through gluon-gluon fusion with a subsequent decay to four $\tau$ leptons.  
%We have attempted to model the triggers at the Tevatron to get a handle on the efficiencies for different decay routes of the four tau leptons.  
By an integrated luminosity of $6 \fb^{-1}$, there will likely be an excess of leptonic events in several different channels at the Tevatron.  
%In  Section \ref{Sec: bba} we present a correlated signal that will be present if this light axion exists -- Yukawa production of the light axion off a bottom quark.   We perform an elementary study and find that the Tevatron's reach should extend beyond the existing limits from LEP2.  
%We conclude with a few comments regarding the future visibility of such 
%a scenario at the LHC.

\section{A Light Electroweak Axion}
\label{Sec: Axion}

In theories with extended Higgs sectors, the masses of the Higgs 
boson fields are expected to be comparable, of order the weak scale. 
However, if there are approximate
$U(1)$ symmetries in the Higgs potential, there can be light pseudo-scalar states after electroweak symmetry breaking.  We call these states axions after 
the best known example of this phenomenon, the original Peccei-Quinn axion.
%, introduced to solve the strong CP problem.  In that case, the $U(1)$ symmetry  was only broken by QCD instanton effects, and led to a sub-eV mass.  
This is merely an example of a more general phenomenon, and it is quite 
possible that the Higgs sector may have an approximate $U(1)$ symmetry that 
gives rise to an axion.

The axions we will discuss here have their symmetries broken more strongly than the QCD axion.  The dominant contribution to the symmetry breaking comes 
not from QCD, but from explicit breaking terms in the Higgs potential.  This large breaking gives the axion a mass in the few GeV range.  If the mass of the $a^0$ is greater than $2 m_{B}\sim 10 \GeV$, then it will
primarily decay into $B$ mesons.  This possibility has been investigated experimentally
at LEP\cite{OpalCascade,DelphiCascade}, and has recently been revisited 
by \cite{Dermisek} in the context of the NMSSM, and \cite{Neal} in more 
general supersymmetric theories.   
Because of the plethora of bottom jets at hadron colliders,  this mode is difficult to study
without detailed Monte Carlo studies, a full exploration of jet clustering algorithms, and an 
understanding of the intricacies of real-world b-tagging.   
We therefore limit our discussion to 
the lighter mass range for the axion where it
is just beneath the bottom quark threshold:  
\begin{eqnarray}
2 m_\tau < m_{a^0} < 2 m_{B}.
\end{eqnarray}
In this case the axions will dominantly decay  to $\tau$ pairs.
This mass range provides a distinctive signal to analyze, especially when some of the taus decay to leptons.  The presence of multiple neutrinos in the final state from tau decays makes reconstruction of  
a mass peak challenging at best.  Nonetheless, an excess of distinctive events coming from Higgs decays 
could be the first hint of beyond the Standard Model physics at colliders.

The presence of an $a^0$ allows for Higgs decays of the form $h^0 \rightarrow a^0 a^0$, with the $a^0$ subsequently decaying to fermions.  One might worry that the presence of a coupling
of the axion to the Higgs might spoil the symmetry keeping the axion light; however,
the Higgs coupling to the axion is derivative in nature, and does not give a large mass to the axion.  
The coupling is given by
\begin{eqnarray}
\frac{c_{haa}}{s} h^0 \partial_\mu a^0\partial^\mu a^0
\end{eqnarray}
where $c_{haa}$ is an $\OO(1)$ parameter dependent number, and the dimensionful quantity $s$
can be interpreted as the axion decay constant.   If the axion comes from an extended Higgs sector, a typical
value is $s\sim \OO(100 \GeV)$.  
This coupling is not suppressed by any small parameters and can overwhelm
the Standard Model $h^0\rightarrow b\bar{b}$ fraction.
If this is a significant branching fraction, then it is possible to look for
Higgs production through gluon--gluon fusion:
\begin{equation}
\label{Eq: Higgs4Tau}
gg \rightarrow h^0 \rightarrow a^0a^0 \rightarrow 4 \tau.
\end{equation}
%%
%If this decay chain dominates, different strategies may be needed to see the Higgs at present and future colliders.

If the Higgs boson dominantly decays in this fashion, it is unclear what the lower bound on its mass is. 
While there is no reason to believe that the model-independent limit
is substantially less than the Standard Model Higgs bound of 114.4 GeV, the highest published bound appears to be 86 GeV\cite{Opal2}.
Recently, there has been some attention to this possibility, emphasizing the fact that alternative decay modes may allow the Higgs boson to be much lighter than the 114.4 GeV bound on a Standard Model-like Higgs boson \cite{Dermisek}, and speculation on whether this might alleviate fine-tuning problems with supersymmetry (see however \cite{Schuster:2005id}).  
%There has even been 
%some discussion of how best to observe this mode at the LHC \cite{GunionLHC}.  
%In Section \ref{Sec: Four tau} we consider observable signals from this mode at the Tevatron.

\subsection{NMSSM Couplings}
\label{Sec: NMSSM}

The search for a Higgs boson decaying into four taus is not strongly dependent on 
many details of TeV-scale physics.  For concreteness, we will detail what is arguably 
the best motivated of such models, but this signal is essentially model-independent.

The supersymmetric Standard Model is the leading candidate for new physics, and
while the MSSM does not allow the Higgs decaying into four taus as a channel,
the simplest extension of the MSSM, the NMSSM,  does.  The NMSSM extends the 
MSSM by adding a gauge-singlet chiral superfield, $S$, to
the TeV-scale spectrum.  There are several different motivations for this extension.
First, it can offer a dynamical explanation for the size of the $\mu$-term.
%In addition, the MSSM $B$-term changes from being a supersymmetry violating term
%to being supersymmetric. 
Finally, it gives an additional contribution to the quartic coupling of the Higgs boson that can increase its mass without having to rely on supersymmetry violating radiative
corrections.  This gives some hope that the fine-tuning present in the MSSM could be ameliorated.

The  superpotential for the NMSSM Higgs sector is given by
\begin{equation}
W= \lambda H_{u} H_{d} S + \frac{\kappa}{3} S^{3}
\end{equation}
There are soft terms as well
\begin{eqnarray}
\nonumber
\LL_{\text{soft}} &=& m^2_u |h_u|^2 + m^2_d |h_d|^2 + m^2_s |s|^2\\
&& + 
(a_\lambda h_u h_d s + \frac{a_\kappa}{3} s^3  +\hc) .
\end{eqnarray}

There are two different limits where a pseudo-scalar becomes light due to
an accidental symmetry of the theory.  The first is the R-symmetric limit \cite{NMSSMR}
where the soft supersymmetry breaking tri-linear couplings vanish
\begin{eqnarray}
a_{\lambda}, a_{\kappa} \rightarrow 0 \hspace{0.3in} \text{R-axion}.
\end{eqnarray}
An R-symmetry does not commute with supersymmetry; therefore,
the R-charges depend on which component of a supermultiplet the field comes from.  For the lowest 
component of the supermultiplets, the R charges for the matter fields ($Q$, $U^{c}$, $D^{c}$, $L$ ,$E^{c}$) and the Higgs fields ($H_{u}$, $H_{d}$, $S$) can all be taken to be R=2/3.  The R charge of gauginos is +1.
The vevs of the Higgs fields spontaneously 
break this symmetry, and if the symmetry were exact, this would lead 
to a massless Goldstone boson.  Gaugino masses and scalar tri-linear $A$-terms violate this symmetry, and give the axion a small mass.

In fact, in the R-symmetric limit, it is difficult to make the
axion mass smaller than a few GeV.  This is precisely because 
the R-symmetry that protects the pseudo-scalar mass is broken by 
the gaugino masses.  $A$-terms are
generated radiatively off of the gaugino masses, and while they can be smaller
than other soft supersymmetry breaking terms, they should not vanish.   
The axion mass is proportional to the generated $A$-terms, and 
cannot naturally be more than a loop factor lighter than the gauginos.  
Limits on charginos from LEP2 then imply:  $m_{a^0}\gsim 1 \GeV.$  
This limit might well be realized, contingent on the specific mechanism of supersymmetry 
breaking.  For example, in gauge mediated and gaugino mediated models the $A$-terms are
generated radiatively from gaugino masses and therefore are smaller
than other soft supersymmetry breaking terms.  It should be noted that there 
are potential issues with the cosmology of the electroweak phase transition in this case \cite{Schuster:2005id}, and the NMSSM has difficulties achieving electroweak symmetry breaking in the simplest models of gauge 
mediation, see e.g. \cite{MurayamaGM}. 

The other limit with a naturally light axion is called the Peccei-Quinn limit\cite{NMSSMPQ},
%, though it should not be confused with the original Peccei-Quinn axion:
%In the PQ limit, the $S$ couplings are small
%%
%\begin{eqnarray}
$\kappa, a_{\kappa} \rightarrow 0$.  % \hspace{0.3in} \text{PQ-axion}.
%\end{eqnarray}
%%
%The Peccei-Quinn charges commute with supersymmetry and are
%%
%\begin{eqnarray}
%Q^{\text{PQ}}_{\text{SM} } = -\half,\hspace{0.3in}
%Q^{\text{PQ}}_{H_u, H_d} = +1, \hspace{0.3in}
%Q^{\text{PQ}}_S = -2 .
%\end{eqnarray}
%%
The PQ limit can arise in $E_6$ or $SU(3)^3$ inspired models which have $\OO(1)$
 $S H_u H_d$ couplings but have small $S^3$ couplings.   
% The PQ limit, unlike the R-symmetric limit, becomes exact as these tri-linear couplings vanish
% and therefore the axion can become arbitrarily light and the axion could 
% in principle be beneath the tau threshold.
%
%In the PQ-axion limit, the axion can be made arbitrarily light,  up to mild phenomenological 
%constraints \cite{Hall:2004qd}. In this limit, however, there is another light state, the 
%fermionic component of the $S$ field, the singlino, $\tilde{s}$. Then decays of the 
%type $h \rightarrow \tilde{s} \tilde{s}$, can be competitive with the $h \rightarrow {a a}$ decays we have discussed.  In this case, a substantial proportion of the Higgs decays can be invisble.

%There is one more way to achieve a light pseudo-scalar axion-like state.  It is of course possible to make the axion light by canceling the various contributions to the mass against one another.  It seems difficult to motivate such a tuning.

The couplings of the $a^0$ to the Standard Model fermions arise through
the mixing of the singlet scalars with the MSSM Higgs with a mixing angle $\theta_{aA}$.  The $a^0$ couples to Standard Model fermions as:
\begin{eqnarray}
g_{aff}\bar{f}\gamma_5 f a^0
\end{eqnarray}
with $g_{aff}$ given by
\begin{eqnarray}
\nonumber
&&\hspace{-0.3in}\frac{- m_f}{v} \cot \beta\cos \theta_{aA}  \hspace{0.2in} \text{(up-type)} \\
&& \hspace{-0.3in}\frac{- m_f}{v} \tan \beta\cos\theta_{aA} \hspace{0.2in} \text{(down-type/leptons)} \label{downtype}
\end{eqnarray} 
with $v= 246\GeV$.  The axion couples like the Higgs, favoring decays to heavy fermions.  The $\tan \beta$ present in the coupling to down-type fermions favors decays to down-type fermions.

The $\tan \beta$  enhancement of Eqn.~(\ref{downtype}) naively points towards a large rate for $a^{0}$ production in association with $b$ quarks.  However, for many regions of the parameter space, the mixing $\cos{\theta_{aA}}$ goes like $1/\tan {\beta}$, spoiling the naive enhancement. This fact precludes the possibility of getting a large enhancement for the $b\bar{b}a$ coupling in the most theoretically motivated regions of the NMSSM.  To see this, note that the usual $\mu$-term of the MSSM can be identified with 
\begin{equation}
\lambda \langle s\rangle \equiv \mu,
\end{equation}
where $\langle s \rangle$ represents the vacuum expectation value of the 
scalar component of the $S$ superfield.  Bounds from chargino searches from 
LEP favor $\mu \gsim 100$ GeV.  Avoiding a Landau pole limits $\lambda \lsim 1$, which means $s$ cannot 
be too small, and in the limit where 
$\langle s \rangle \tan \beta \gg 2 v$, the mixing angle is given by 
\begin{equation}
\cos \theta_{aA} \approx \frac{2 v}{s \tan \beta}  \hspace{0.5in} \rm{R-axion}
\end{equation}
with $v = 246 \GeV$.  Then, the coupling to $b$ quarks is given by roughly $\frac{2 m_{b}}{s}$. 

\subsection{Existing limits}
\label{Sec: Existing Limits}

One might worry that such a light $a^0$, $m_a < 2 m_B$, would have already been 
excluded or discovered by existing experiments.  This is not the case.

B-physics experiments at $b$ factories looked for (strong CP) axions via the process $\Upsilon \rightarrow \gamma \, +$ invisible.  However, 
the axion considered here will decay, so these limits do not directly apply. An applicable but weaker limit arises from searches for monochromatic photons, with the axion decaying to any charged Standard Model particles \cite{CLEOCUSB}.  However, these bounds become very weak as $m_{a}$ approaches 2$m_{b}$. 
Limits from LEP on the processes of Eqn.~(\ref{Eq: Higgs4Tau}) and Eqn.~(\ref{Eq: Axionstrahlung}) exist, but 
do not exclude this scenario for all couplings.  

The strongest limit on this process appears to come from two OPAL papers.  In \cite{Opal1}, the process $(e^{+} e^{-} \rightarrow b \bar{b} a\rightarrow b \bar{b} \tau \tau)$ was searched 
for.   In the region  4 GeV $< m_{a} <$ 10 GeV, the $a^0 b \bar{b}$  coupling $y_{abb}$ is constrained, 
\begin{eqnarray}
y_{abb} \lsim 10 \frac{m_{b}}{v}.
\end{eqnarray}
In \cite{Opal2}, limits are placed via the process $(e^{+} e^{-} \rightarrow Z^{*} \rightarrow Z h \rightarrow Z a a \rightarrow Z 4\tau)$.  Strong limits exist in the region $m_{h} < 86$ GeV.  In principle, existing data 
might be able to extend them (as stressed in \cite{Neal}). The analysis only cuts off because the region is ``theoretically disfavored in the MSSM.''

\section{Leptonic Excesses at the Tevatron}

The relatively weak limits on the Higgs boson decaying to four taus from LEP2 present an opportunity for the Tevatron to extend the search.   The signal is distinctive enough that cuts can implemented to get away from the backgrounds present at a hadron collider.  The kinematics of the events are important in designing these cuts.  We are going to consider two different production mechanisms for the Higgs: gluon-gluon fusion and associated production with a weak vector boson.

\subsection{Gluon Fusion}

Gluon fusion is the dominant production mechanism of the Standard Model Higgs boson at the Tevatron.
With a production cross section of nearly 1 pb (see, e.g., \cite{Spira}),  there may be thousands of Higgs produced at the Tevatron.  We wish to investigate whether this scenario might be usefully probed at the Tevatron before the LHC even turns on.

For a Standard Model-like Higgs, this channel is hopeless, as it is impossible to reconstruct the $b \bar{b}$ mass peak above the enormous QCD background.  If  Higgs decays dominantly end in a four tau final state, it is possible to harness 
this large cross section.   

The size of the gluon fusion cross section is controlled by the coupling of the Higgs to top quarks.  In the R-axion limit of the NMSSM,  the Higgs production cross section is within a few percent of the Standard Model's, and may be large enough to allow discovery with sufficient luminosity.  As usual, the associated vector boson production is smaller.  It appears unlikely that an initial discovery would be made in this channel. After the branching ratios of the four taus are folded into the small production cross section, there will not be many events. Unlike the Standard Model case, gluon fusion is the most promising channel; this is entirely due to the distinctive nature of the signal.

\subsubsection*{Basic Kinematics}
We have in mind a Higgs boson with mass $m_{h} \sim 100$ GeV, and $m_{a} \sim 8$ GeV.  At the Tevatron,  Higgs bosons are produced via gluon-gluon fusion centrally.  The Higgs boson decay will give rise to two back-to-back $a^{0}$'s each with a large boost factor.  As a result, the decay products of the $a^{0}$ will be collimated.  To gain a basic understanding of the efficiencies of the signals discussed in this section, we used a modified version of Pythia\cite{Pythia} to generate signal events, which we then piped through Tauola\cite{Tauola} and then through PGS, a simple detector simulator\cite{PGS}, using its standard implementation of the CDF detector.  

Efficiencies for leptons will be crucial to the observability of the signal.  We imposed the standard PGS isolation cuts for the electrons with one minor modification.  The isolation cuts limit the amount of nearby energy in the calorimeter and the number of tracks within an isolation cone of R=0.4.  We slightly relaxed this cut to allow the presence of a muon within that cone.  The boost of the $a_{0}$ often leads signal events to have this topology.  For muons, the only isolation cut imposed was that it not lie within a $\Delta R= 0.7$ cone of a jet.   The detector simulation has electron and muon coverage out to an $\eta=2.0$.

There are several factors that make observing this signature challenging.  First, the Higgs bosons
are predominantly produced with small $p_T$.  Their decay particles
will be isotropically distributed throughout the detector, and with four-body
decays the probability that some number of particles are lost is non-negligible.  
Second, in leptonic decays of the $\tau$s, a large fraction of the energy is carried away by the 
neutrinos, resulting in soft leptons.  

\begin{figure}[h!]
\begin{center}
\includegraphics[width=\columnwidth]{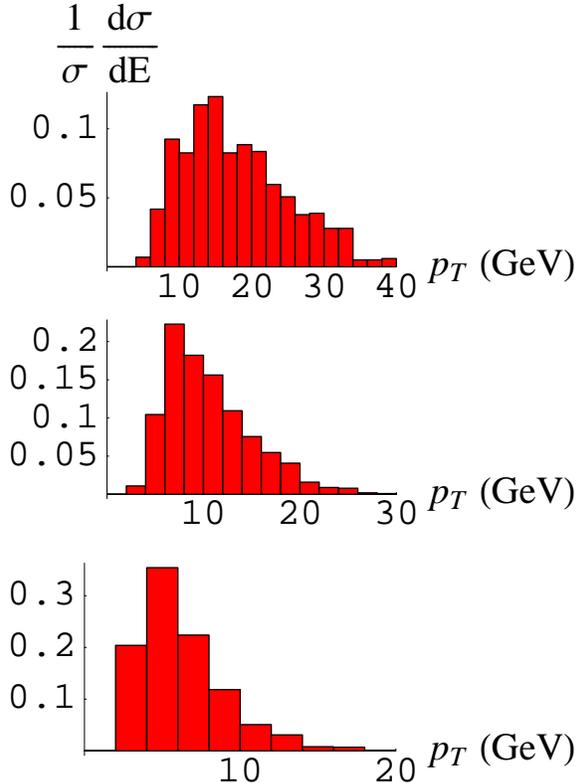}
\caption{
\label{Fig: LeptonPT}
The $p_T$ distribution of the leptons in tri-lepton events coming from $h^0\rightarrow a^0a^0\rightarrow 4\tau$.  We have imposed cuts of $\eta_\ell < 2.0$, and $p_{T}^{\ell} > 3.0$ GeV, as well as rudimentary isolation cuts as described in the text. 
}
\end{center}
\end{figure}

The spectra of the three hardest leptons are plotted in Fig. \ref{Fig: LeptonPT}.  As alluded to 
earlier, the softness of the lepton spectra should be accounted for when optimising cuts.  

\subsubsection*{Tri-Leptons}

One channel with very little Standard Model background is the case where at 
least three of the final state $\tau$'s decay leptonically.  Then we have a 
tri-lepton signature, a final state well known from supersymmetry  searches at the Tevatron\cite{Run2Sugra}.  As shown in that context, the Standard Model background 
can be made tiny ($<$ 1 event/fb$^{-1}$) with an appropriate set of cuts.
There are two relevant caveats.  First, the three leptons will be somewhat softer than those expected  from the traditional supersymmetry signature.   Since 
the Higgs boson mass is shared between the observed leptons and several neutrinos, 
the leptons will have energies of roughly 10 GeV.   Because of this, care must be taken to set hardness cuts  appropriately, while avoiding the background from soft leptons coming from off-shell photons. 

Since these events contain a substantial number of neutrinos, it is impossible to reconstruct the 
Higgs boson mass.  In fact, they should be marked by the presence of a rather substantial amount of 
missing energy.  However, this is unlikely  to be of much assistance in decreasing the background.  
One significant Standard Model background comes from di-boson 
production ($Z^0W^\pm$ or $\gamma W^\pm$), which also has large missing energy from the neutrino in the $W^\pm$ decay. 

The tri-lepton Higgs decays make up 12.7\% of the total branching fraction of 
four $\tau$ events.  There are two challenges in observing this signal: the third lepton
typically is soft, and all three leptons must lie within the geometric acceptance
of the detector.  The $p_{T}$ distribution of the three leptons in tri-lepton events are shown in the Fig. \ref{Fig: LeptonPT}.  Using PGS, we estimated the efficiency, $\epsilon$ times branching fraction $B$ for tri-lepton events coming from Higgs bosons decaying to four taus.  We find $\epsilon B \sim 0.5 \%$, corresponding to 
\begin{eqnarray}
&& \epsilon B \sigma(gg \rightarrow h \rightarrow 3 \ell)=4-6.5 \text{ fb} \nonumber \\
&& (120 > m_{h} > 100 \text{ GeV}) 
\end{eqnarray}
This leaves very little room for additional cuts.  These channels fortunately have very few Standard Model backgrounds.  The main physics background should be dominated by events with two weak vector bosons: $\gamma^*\gamma^*, Z^0{}^*\gamma^*, Z^0 W^\pm$ production.

\subsubsection*{Opposite Flavor Di-lepton Signal}

A promising channel for finding four $\tau$ events is two $\tau$s decaying to opposite flavor di-leptons.
Requiring opposite flavors requires the presence of either two leptonically decaying 
taus, or a double weak process, so this already substantially reduces the potential background.  In the case where both leptons arise from the decay of the same $a_0$, the two leptons will be well-collimated.  Since there are many Standard Model backgrounds that produce back-to-back $\tau$'s, we restrict ourselves to looking for same-side di-leptons.  While the invariant mass of the di-leptons is small (less than 10 GeV, due to the light $a^0$), the summed 
$p_T$ need not be nearly as small.  In Figure 2 we show the distribution of $\Delta R \equiv \sqrt{\Delta \phi^{2} + \Delta \eta^{2}}$ between the same sign leptons.  This distribution shows that care must be taken in the implementation of the electron isolation cut to avoid cutting away the signal.  

\begin{figure} 
\label{Fig:DeltaR}
\includegraphics[width=3.0in]{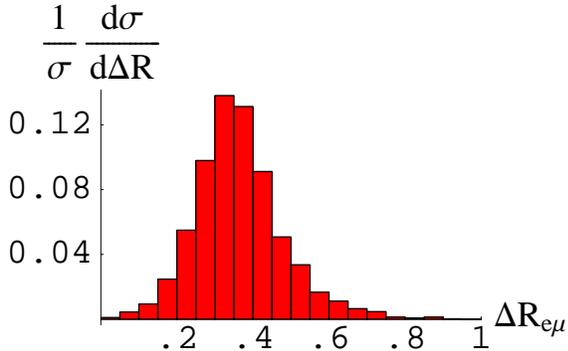}
\caption{The $\Delta R \equiv \sqrt{ \Delta \phi^{2} + \Delta \eta ^{2}}$ of the e-$\mu$ pair.  Note that a substantial portion of the signal lies in the region $\Delta R < 0.4$, which might be eliminated by a naive isolation cut.  The opening angle is controlled by the ratio $m_a/m_{h}$.  Here, $m_{h}=100$ GeV and $m_{a^{0}}$=8 GeV. }
\end{figure}

One Standard Model background arises from neutral currents to $\tau$ pairs, with the low 
invariant mass making this a large background.  For the vast majority of this 
background, however, the resulting $e\mu$ pair will always be on opposite sides
of the detector, i.e. $\Delta \phi_{e\mu} > \frac{\pi}{2}$ because the neutral current resonance will
not have a significant $p_T$.  Meanwhile, the signal will have a peaked distribution 
of $\Delta R_{e\mu}$  with many having a small opening angle $\Delta R < 0.4$.  
Folding in the branching ratios for the tau decays, and noting that an $e-\mu$ pair coming from the same $a^{0}$ will have a small $\Delta \phi$, roughly 11.8\% of all 
$h^0\rightarrow 4 \tau$ will be opposite flavor same side events. This leads to a signal cross section of $\sim 100$ fb before any efficiencies are applied.

While requiring the OFDL to be on the same side reduces many Standard Model backgrounds considerably, the cross section for jet plus neutral current is not negligibly small, and the $p_T$ of the 
jet can balance the $p_{T}$ of the neutral current resonance, leading to a small opening 
angle for the lepton pair, just like the signal.   Since we are requiring opposite flavor di-leptons, to mimic the signal the neutral current must decay into a pair of taus and subsequently to leptons.  The invariant masses of $a^0\rightarrow e\mu \MET$ will be in the 4 to 10 GeV range.  The most important resonance in this range is $\Upsilon(1S)$.  The
$p_T$ of the $e\mu$ pairs in the $h^0\rightarrow e\mu +X$ will be $\gsim 20$ GeV, which corresponds
to a $p_T$ of the $a^0$ or $\Upsilon$ of $\gsim 40$ GeV (due to the missing neutrinos).  The $p_{T}$ distribution of the same-sign dilepton pairs is shown in Figure 3.
The cross section for  $\Upsilon + X$  is a steeply falling function of $p_T$.  Extrapolating the measurement of the CDF cross-section \cite{Cropp:1999ub} for upsilon production to higher $p_T \sim 40$ GeV, we estimate the cross-section for $\Upsilon$ production with such a large $p_{T}$ to be $\sigma \sim 10$ fb, which could in principle compete with the signal.   Extending the measurement of $\Upsilon$ production out to larger $p_{T}$ will be an important part of extracting this signal.  A flavor subtraction using the $\Upsilon \rightarrow \mu \mu$ channel might also be possible.

\begin{figure}
\label{Fig:emuptdist}
\includegraphics[width=3.0in]{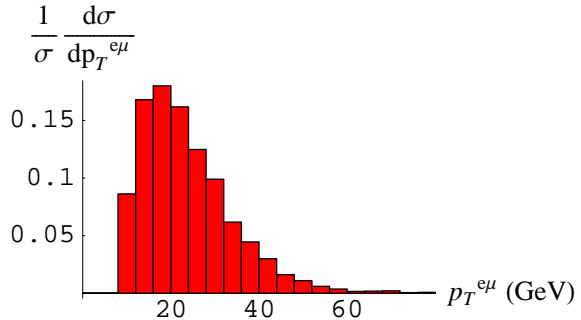}
\caption{The $p_T$ distribution of the close e-$\mu$ pair.  Events are selected where there is a e-$\mu$ pair with a $\Delta R < 1.0$.}
\end{figure}

 There is another important distinction between $\Upsilon$ production, and the signal.  A high $p_{T}$ $\Upsilon$  most often recoils against a high $p_{T}$ gluon jet, while the leptons from the signal are recoiling against a pair of $\tau$s. A high $p_T$ jet looks quite different from a pair of (overlapping) taus and thus can be used to discriminate signal from background. The high $p_T$ jet will typically have many tracks.  This is in contrast to the  four $\tau$ events that will typically only have a few additional tracks (arising from the $\tau$ decays on the opposite side) in addition to the $e\mu$.   So, if we also require that
these same-side, opposite flavor events are ``quiet'', i.e. do not have a large number of 
high $p_T$ tracks, then this forms another
useful discriminant. The events that have same-side opposite flavor leptons can either have two hadronic taus on the opposite side, with typically 6 tracks, a 
hadronic and leptonic tau with typically 4 tracks or occasionally a doubly leptonic event with 
only a few tracks. 

\begin{figure}[htb]
\begin{center}
\includegraphics[width=3.0in]{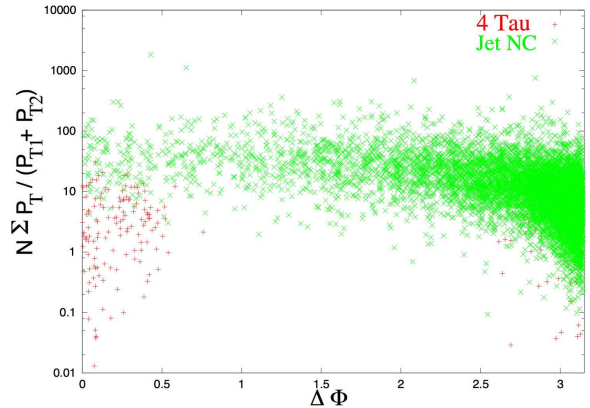}
\caption{
\label{Fig: OF Discriminant}
One possible discriminant between opposite flavor, opposite charge di-lepton events and $j \tau\tau$ events.  The discrimant, $M$, described in the text versus  $\Delta \phi_{e\mu}$.  Notice that the four tau opposite flavor events are clustered at small $M$ close to $\Delta \phi_{e\mu} =0, \pi$ whereas the jet di-tau events have much higher $M$'s, thus providing a clean discriminant between the two types of events.   
}
\end{center}
\end{figure}

A simple way to quantify the differences in the events between signal (4 $\tau$) and background 
$(\text{NC}+j)$ is by defining a ``quiet'' event.   
 In order to balance the $p_T$  of opposite flavor jets on the same side of the detector, there needs to be a high $p_T$ jet.  
These jets will typically have large multiplicities in contrast to four $\tau$ events that
will only have at most two hadronic taus to balance the $p_T$ against.  A criterion that distinguishes the particle multiplicity on the opposite side of the event should differentiate a portion of the background events from the signal.  An example of such a discriminant is given by 
\begin{eqnarray}
M \equiv  \frac{N_p  \sum E_T}{p_{T\, \ell_1} + p_{T\, \ell_2}} .
\end{eqnarray}
where $N_p$ is the number of particles with $E_T > 0.5 \GeV$ (in order to avoid counting too many particles from underlying events) and $|\eta|< 3.0$ excluding the opposite flavor leptons,  $\sum p_T$ is the sum of the momentum of those $N_p$ particles and $p_{T\, \ell}$ is the $p_T$ of each of the leptons.   The neutral current plus jet background will have large $M$, while the signal will have a smaller $M$.   We simulated jet plus neutral current and four tau events in Pythia and found that it was possible to get a rejection rate of greater than $10^3$ while losing only one third of the signal.  This is shown in Fig. \ref{Fig: OF Discriminant}.
The use of this variable gives hope that one might bring the 
neutral current plus jet background under control.   Simple Monte Carlo simulations are not entirely reliable in this type of study, but this simple example makes it seem plausible that such a discriminant could be designed.

The most serious background arises from heavy flavor jets that undergo double semi-leptonic decays where the jet is lost.  This background has to be measured in order to get a handle
on its size.  Fortunately this is possible.  By requiring a good b-tagged jet on one side of the event and looking for di-leptons with no jet on the other side, one can estimate this background.

Since heavy flavor events are also strongly produced, it will again be possible to place a quiet cut on the balancing jet of the event in order to reduce the background.  We did not model the quiet cut in this case,  
but if the cut is as efficient as for the neutral current plus jet events, then a b-jet can 
fake a pair of leptons at the rate of $10^{-5}$ without completely swamping the signal.

\subsubsection*{Other Channels}

Since they make up the dominant decay, it would be ideal if hadronic decays of the tau could be used.  There are two important caveats.  The first is that two taus often overlap in the signal.  If a hadronic tau decay overlaps with a leptonic decay, the semi-leptonic decays of heavy jets could again be a serious background.  In the case where two hadronic taus overlap, the standard tau identification cuts would clearly fail, if for no other reason than that two ``one-prong'' decays would now appear as a jet with two tracks.  In this case, the dominant backgrounds again appear to come from jet flucutations.  While there might potentially be thousands of signal events in hadronic tau channels at the Tevatron, it is difficult to estimate how many might survive the draconian cuts that would have to be applied to get away from the jet backgrounds.   Finally, it might also be possible in principle to search for same-sign leptons, but these two leptons would be coming from opposite sides of the event, and unless a third tau decayed leptonically, each lepton would be close to hadronic activity, and one would again worry about heavy flavor backgrounds.

\subsection{Associated Production}
Higgs bosons are also produced at the Tevatron in association with gauge bosons.  The cross sections for these processes are substantially smaller than the gluon fusion production cross section, and as a result, are less likely to be the first source of anomalous events.  Nevertheless, it is possible that these events can still contribute to the interesting final states discussed above.  The first possibility is that a combination of leptons from the gauge boson decays and the Higgs boson decay can lead to tri-lepton (or four lepton) final states.  Unless there is a lepton coming from the gauge boson decay, this contribution is expected to be subdominant to the tri-lepton signature coming from $gg \rightarrow H \rightarrow 4 \tau$.  The reason is simply that the diminished cross-section\cite{Spira} 
\begin{eqnarray}
\nonumber
&&\sigma(p \bar{p} \rightarrow W H)  \approx 200-400 \fb\\ 
\nonumber
&& \hspace{0.8in} (120 > m_{h} > 100 \text{ GeV}) \\
\nonumber
&&\sigma(p \bar{p} \rightarrow Z H)  \approx 100-200\fb\\
&& \hspace{0.8in} (120 > m_{h} > 100 \text{ GeV}),
\end{eqnarray}
is not overcome by a boost in acceptance.  The sole difference in acceptance from the gluon fusion case results from the boost that the Higgs gets from recoiling against the gauge boson.  While this can lead to harder leptons, it does not overcome the small cross section.  Using PGS, we estimated the efficiency for tri-lepton events where the gauge bosons decayed hadronically.  Taking into account the branching 
ratio, $B$, of $\tau \rightarrow$ leptons, we found an efficiency times $B$ of roughly $\epsilon B=0.6\%$, leading to a contribution to a tri-lepton signal of:
\begin{eqnarray}
\nonumber
&&\hspace{-0.3in}\epsilon B \sigma(\text{trilepton from Higgs, V $\rightarrow$ hadrons}) = 1-2 \fb\\
&&   \hspace{0.8in} 120 > m_{h^0} > 100\GeV
\end{eqnarray}
In the case where the $W$ decays leptonically, the hard lepton from that decay might give rise to an additional contribution to the tri-lepton signal.  Again, using PGS, and taking into account the tau branching fractions, we estimate:
\begin{eqnarray}
\nonumber
&&\hspace{-0.3in}\epsilon B \sigma (\text{2 lepton from Higgs, V $\rightarrow$ lepton(s)}) = 3-5\fb\\
&& \hspace{0.8in} 120 > m_{h^0} > 100\GeV 
\end{eqnarray}
The leptonic decays of the $Z$ are unlikely to be helpful in generating a distinctive tri-lepton signature, as $WZ$ backgrounds from the Standard Model may be an issue.  On the other hand, these decays might occasionally lead to rather spectacular four-lepton events, but the rate of such events is quite small ($\lsim$ 0.5 event / fb$^{-1}$).

Finally, there is a question of whether one might look for events with the distinctive close $e \mu$ pair in associated production events.  The largest number of such events comes in the case where the gauge boson decays hadronically.  Taking into account the branching ratio for the $\tau$'s to decay to the appropriate leptons, a PGS simulation yielded an efficiency for these events of roughly 2\%.  However, to avoid the backgrounds arising from $\Upsilon+X$, it would be desirable to impose that the jets reconstruct the gauge boson mass.  This further reduces the efficiency by approximately a factor of two.  The result is a rate of 
\begin{equation}
\epsilon B \sigma (V+ (e \mu)) = 3-7\fb  \hspace{0.3in} 120 > m_{h^0} > 100 \text{ GeV}  
\end{equation}
This is unlikely to be observable above the $\Upsilon + j$ background, as it seems difficult to impose a ``quiet cut'' on jet activity in this case.      

The small rates for the associated production processes make observation difficult.  We conclude that these processes may give a boost to the tri-lepton signature, but are unlikely to be otherwise observable.

\section{Conclusion}
Because the dominant backgrounds to the signals discussed here are likely to come from heavy quark decays faking leptons, it is not possible to calculate whether the signal will be visible above background.  However, given the prospect of thousands of Higgses decaying in this way at the Tevatron, it is worthwhile to measure the backgrounds, and look for these distinctive decays.

If an excess of events are seen in one of the channels that have been described here, it will be difficult to definitively determine that this decay chain is the underlying cause.  The best hope is to observe multiple excesses.  An observation of the distinctive $e-\mu$ topology described here, along with an excess of multi-lepton events, would go a long way towards indicating the presence of a low-mass state that couples to $\tau$ leptons.  If information from the hadronic tau decays can be harnessed, the case for new physics in taus will be even more compelling.  Another hint might come from searching for the existence of the light tau-philic final state:
\begin{equation}
\label{Eq: Axionstrahlung}
p\bar{p} \rightarrow b \, \bar{b} \, a^0 \rightarrow b \bar{b} \tau^{+} \tau^{-}
\end{equation}
If at least one $\tau$ decays leptonically and there is at least one $b$-tagged jet,then this might be visible above background,  though backgrounds from leptonically decaying b-jets seem particularly dangerous.  Nevertheless, at a minimum it seems the Tevatron might well improve the bounds on this process from LEP\cite{Opal1}.

At the LHC, the number of Higgs events will be much larger, and so it is worthwhile attempting to reconstruct the mass peak, perhaps through subdominant decays (see, e.g. \cite{GunionLHC}).  This model might also have a distinctive signature in events where the $a^{0}$ is radiated off of a top quark ($t$-$\bar{t}$-$a^{0}$).  While the coupling to the top is model-dependent, and may be supressed by $\tan \beta$, this class of events could lead to a very striking decay chain if the tops decay leptonically: $t \bar{t} a^{0} \rightarrow b \bar{b} W W \tau^{+} \tau^{-} \rightarrow b \bar{b} \mu^{+} \mu^{+} e^{-} e^{-}$.  However, ensuring all leptons are sufficiently isolated to be identified may be a challenge.  

The inability to reconstruct a mass peak for the Higgs boson in this channel may make it difficult to unambigiously identify this decay chain as the source of leptonic excesses.  On the other hand, the distinctive nature of the kinematics gives hope.  For example, if back to back e-$\mu$ pairs were seen, each with tiny opening angle and a $p_{T}$ of $\sim 30$ GeV, it would appear to point strongly to this scenario.

One lesson learned from this scenario is that even high $p_{T}$ physics such as Higgs decay can potentially lead to signals with relatively soft leptons whose isolation is imperfect.  This is not a peculiar feature of this model.  While here it is $m_{h} \gg m_{a}$ that leads to the unisolated leptons, 
a similar effect should occur in any situation where there is a hierachy of new physics scales.  If the new particles with small mass have small direct couplings to light Standard Model fields, such models might be missed unless care is taken to search for these event topologies.

\section*{Acknowledgments}

We would like acknowledge Amit Lath, Andy Haas, Henry Frisch, Michelangelo Mangano, Matt Strassler and Scott Thomas for useful conversations, and John Conway and Jesse Thaler for assistance with PGS.  We would like to thank the Theory Group of Rutgers for their hospitality during the initial stages of the work.  JW would like to thank the Theory Group of the University of Texas at Austin for their hospitality during the completion of this work.   The work of PG and JW is supported by National Science Foundation grant PHY-9870115 and the Stanford Institute for Theoretical Physics. 
The work of AP is supported by the DOE under contract DE-FG02-91ER40654.

\end{document}